\documentclass[twocolumn,showpacs,prl]{revtex4}
\usepackage{amsmath}
\usepackage{mathrsfs}
\usepackage{graphicx}
\begin{document}

\draft

\title{Decoherence of the Kondo Singlet via a Quantum Point Contact Detector}%
\author{Kicheon Kang}%
\email{kckang@chonnam.ac.kr}
\affiliation{Department of Physics
and Institute for Condensed Matter Theory, Chonnam National
University,
  Gwangju 500-757, Korea}%

\date{\today}

\begin{abstract}
We investigate the effect of the charge state measurement of the
Kondo singlet for a quantum dot transistor via a capacitively
coupled quantum point contact detector. By employing the
variational ansatz for the singlet ground state of the quantum
dot combined with the density matrix formulation for the coupled
system, we show that the coherent Kondo singlet is destroyed by
the phase-sensitive as well as the magnitude-sensitive detection
in the transmission/reflection coefficients at the quantum point
contact. We argue that the phase-sensitive component of the
decoherence rate may explain the anomalous features observed in a
recent experiment by Avinun-Kalish {\em et al.} (Phys.~Rev.~Lett.
{\bf 92}, 156801 (2004)). We also discuss the correlations of
the shot noise at the quantum point contact detector and the
decoherence in the quantum dot.
\end{abstract}


\maketitle

\let\veps=\varepsilon%
Decoherence induced by the measurement of charge in mesoscopic
electronic devices provides an ideal playground for studying the
wave-particle duality in quantum mechanics. Experiments on the
controlled dephasing were performed in mesoscopic structures based
on quantum dots (QD) in the Coulomb blockade
limit~\cite{buks98,sprinzak00}. While the coherent transmission of
electrons through a QD is monitored by using an Aharonov-Bohm
interferometer~\cite{yacoby95,schuster97}, a nearby quantum point
contact (QPC) capacitively coupled to the QD (weakly) measures
the charge state of the QD, and suppresses quantum coherence.
This measurement-induced decoherence is controlled via the
applied voltage across the QPC detector. Various different
methods have been used to study this problem
theoretically~\cite{aleiner97,levinson97,gurvitz97,stodolsky99,buttiker00}.

Recently, this kind of controlled dephasing experiment was also
performed in the Kondo limit of the QD~\cite{kalish04}. Kondo
singlet is formed between the localized spin in a QD and
electrons in the leads~\cite{hewson93}, which gives rise to
enhanced transport through the
QD~\cite{gordon98,cronen98,schmid98,simmel99,ji00,wiel00,kouwenhoven01}.
It was shown that a nearby QPC capacitively coupled to the QD plays a
role of a ``potential detector" and raises significant
suppression of the Kondo resonance~\cite{kalish04}.
However, characteristics of the
measured suppression were very different from the theoretical
prediction of Ref.~[\onlinecite{silva03}]. The most significant deviation from
the theory is that the measured suppression strength is much
larger (about 30 times) than expected. Dependence on the
transmission probability ($T_d$) and on the bias voltage ($V_d$) across
the QPC were also inconsistent with the theoretical expectation.

The analysis of the experiment~\cite{kalish04} was based on a
theory~\cite{silva03} of suppression of the Kondo resonance due
to the path detection by the QPC via the change of the
transmission {\em probability}, $\Delta T_d$. We point out that
this treatment is incomplete, because scattering of electrons at
the QPC is a quantum mechanical phenomenon with complex
transmission and reflection amplitudes. Therefore {\em
phase-sensitive} detection should also be taken into
account~\cite{sprinzak00,stodolsky99,hacken01}, in general.

In this Letter, we present a theory of controlled dephasing for
the Kondo singlet via charge detection of a nearby QPC. First, we
show that the change of the transmission (and reflection) {\em
phase} as well as its probability change induces decoherence of
the Kondo-correlated state, unlike in
Ref.~\cite{kalish04,silva03}. Second, by using a simple QPC model,
we argue that the phase-sensitive contribution to the decoherence
may be much larger than the magnitude-sensitive component (due to
$\Delta T_d$) in a typical geometry of the QD-QPC hybrid
structure. In addition, we discuss on the $T_d$-dependence of the
decoherence rate which might be correlated with the
characteristics of the shot noise in the QPC.

To describe the Kondo singlet of the QD, we adopt the variational
ground state for the impurity Anderson
model~\cite{hewson93,gunnarsson83}, which is known to describe
the essential Kondo physics in a very simple but effective way.
Further, this approach can be easily combined with the density
matrix formulation in the presence of a detector. The Hamiltonian
is given by
\begin{subequations}
\begin{equation}
H = H_L + H_R + H_D + H_T \;.
\end{equation}
The left (L) and the right (R) leads are described by the noninteracting
Fermi sea as
\label{eq:hamil}
\begin{equation}
H_\alpha = \sum_{k\sigma}\varepsilon_{\alpha k}
c^\dagger_{\alpha k\sigma} c_{\alpha k\sigma} \quad
(\alpha = L, R) \,,
\end{equation}
where $c_{\alpha k\sigma}$ $(c_{\alpha k\sigma}^\dagger)$ is an
annihilation (creation) operator of an electron with energy
$\varepsilon_{\alpha k}$, momentum $k$, and spin $\sigma$ on the
lead $\alpha$.
The interacting QD is described by
\begin{equation}
H_D = \sum_\sigma \varepsilon_d d_\sigma^\dagger d_\sigma
+ U n_\uparrow n_\downarrow  \,,
\end{equation}
where $d_\sigma$ and $d_\sigma^\dag$ are QD electron operators,
$n_\sigma=d_\sigma^\dag d_\sigma$, $\varepsilon_d$ and $U$ stand for the
energy of the localized level and the on-site Coulomb interaction,
respectively.
The tunneling Hamiltonian $H_T$ has the form
\begin{equation}
\label{kondo-phase:1c}
H_T = \sum_{\alpha=L,R}\sum_{k\sigma}
\left(V_\alpha d_\sigma^\dag c_{\alpha k\sigma} + h.c.\right) \;,
  \end{equation}
where $V_\alpha$ is responsible for for the tunneling between the QD and
  the lead $\alpha$.
  \end{subequations}

In the absence of interaction between the QD and the QPC, the variational
ground state for the Hamiltonian $H$ ($U\rightarrow\infty$ limit) is written
as~\cite{gunnarsson83}
\begin{subequations}
\begin{equation}
 |\Psi_G\rangle = A \left( |0\rangle + |1\rangle \right) \;,
\end{equation}
where $|0\rangle$ denotes the Fermi sea for the leads with empty QD state, and
\begin{equation}
   |1\rangle \equiv  \frac{1}{\sqrt{2}} \sum_{\alpha\sigma,k<k_F}
   v_{\alpha k} d_\sigma^\dagger c_{\alpha k\sigma} |0\rangle \,.
\end{equation}
Here $A=\sqrt{1-\bar{n_d}}$, with $\bar{n_d}$ 
being the average occupation
number of the QD level, and $v_{\alpha k} =
\sqrt{2}V_\alpha/(E_G-\varepsilon_d +\varepsilon_{\alpha k})$
where $E_G$ denotes the ground state energy determined by the
equation
\begin{equation}
 E_G = 2\sum_{\alpha,k<k_F}
       \frac{ V_\alpha^2 }{ E_G-\varepsilon_d+\varepsilon_{\alpha k} } \;.
\end{equation}
The Kondo temperature ($T_K$), characteristic energy scale of the
system, is given as a difference between the QD level
($\varepsilon_d$) and the ground state energy ($E_G$):
$T_K=\varepsilon_d-E_G$.
\end{subequations}

In fact the states $|0\rangle$ and $|1\rangle$ have different
occupation numbers for the QD, $n_d=0$ and $n_d=1$, respectively.
A capacitively coupled QPC to the QD is able to detect the charge
state, since the potential of the QPC depends on the charge state
of the QD. So the transmission and reflection amplitudes through
the QPC also depend on $n_d$. To describe this situation, it is
effective to introduce the density matrix
formulation~\cite{gurvitz97,hacken98,hacken01}. We assume that
the QPC supports only a single transverse mode so that the
scattering matrix $S_{\mathsf{QPC}}$ through the QPC is a
$2\times2$ matrix that depends on the charge state of the QD:
\begin{subequations}
\begin{equation}
 S_{\mathsf{QPC}} = \left\{ \begin{array}{cc}
                     S_0 & \mbox{\rm for } |0\rangle , \\
                     S_1 & \mbox{\rm for } |1\rangle ,
           \end{array} \right.
\end{equation}
where
\begin{equation}
 S_\alpha = \left( \begin{array}{cc}
                    r_\alpha & t_\alpha' \\
            t_\alpha & r_\alpha'
           \end{array}  \right) \;\;\; (\alpha=0,1).
\end{equation}

We can describe the coupled system by a two-particle scattering
matrix~\cite{hacken98,hacken01}
\begin{equation}
 S_{\alpha\alpha'} = \delta_{\alpha\alpha'} (\delta_{\alpha 0}S_0
    + \delta_{\alpha 1}S_1 )
\end{equation}
\end{subequations}
where $\alpha,\alpha'\in \{0,1\}$. With $\rho_\mathsf{tot}^0 =
\rho^0\otimes\rho_\mathsf{QPC}^0$ being the density matrix of the
total system before the passage of an electron through the QPC,
the density matrix after scattering  is given by
$\rho_\mathsf{tot} = S \rho_\mathsf{tot}^0 S^\dagger$. The
reduced density matrix of the QD is obtained by tracing out
$\rho_\mathsf{tot}$ over the QPC degree of freedom as
\begin{subequations}
\begin{equation}
 \rho = \mathsf{Tr}_\mathsf{QPC} \rho_\mathsf{tot} \;.
\end{equation}
It is found that the diagonal elements of $\rho$ do not change
upon scattering through the QPC, but the off-diagonal elements are
modified by
\begin{equation}
 \rho_{01} = \lambda\rho_{01}^0  ,
\end{equation}
where
\begin{equation}
 \lambda = r_0r_1^* + t_0t_1^* \,. \label{eq:lambda}
\end{equation}
The QD-charge-dependent transmission/reflection amplitudes are
complex numbers and can be rewritten as ($\alpha=0,1$ denoting
the charge state of the QD)
\begin{eqnarray}
 t_\alpha &=& |t_\alpha|\exp{(i\phi_{t\alpha})} \;, \\
 r_\alpha &=& |r_\alpha|\exp{(i\phi_{r\alpha})} \;,
\end{eqnarray}
\end{subequations}
satisfying the relation $ |t_\alpha|^2+ |r_\alpha|^2=1$.

We consider the limit where the scattering through the QPC takes place
on a time scale much shorter than the relevant time scales in the QD.
In our case, $\Delta t \ll t_d$, where $\Delta t=h/2eV_d$ denotes the
average time between two successive scattering events with $V_d$ 
being the voltage across the detector, and $t_d$ is the
decoherence time of the QD. In this limit one finds that
\begin{subequations}
\begin{equation}
 \rho_{01}(t) = e^{ (i\Delta\epsilon-\Gamma_d) t }
 \rho_{01}(0) \,,
\end{equation}
where
\begin{eqnarray}
 \Delta\epsilon  &=& \frac{1}{\Delta t}\arg{\lambda} \,, \\
 \Gamma_d &=& \frac{1}{t_d} = -\frac{1}{\Delta t} \log|\lambda| \;. \label{eq:1/td}
\end{eqnarray}
\end{subequations}
On the other hand, the diagonal terms are independent of time,
which implies that no relaxation of the charge state takes place.

In the weak measurement limit ($\lambda\approx1$), both $\Gamma_d$ and
$\Delta\epsilon$
can be obtained in terms of the change in the magnitude and phase
of transmission/reflection amplitude of the QPC:
\begin{subequations}
\begin{equation}
 \Gamma_d = \Gamma_T + \Gamma_\phi
\end{equation}
where
\begin{eqnarray}
 \Gamma_T &=& \frac{eV_d}{h} \frac{(\Delta T_d)^2}{4T_d(1-T_d)} \;, \\
 \Gamma_\phi &=& \frac{eV_d}{h} T_d(1-T_d) (\Delta\phi)^2 \;,
\end{eqnarray}
 \label{eq:1/td-weak} \\
\end{subequations}
and
\begin{equation}
 \Delta\epsilon = \frac{eV_d}{\pi}(1-T_d)\Delta\phi_r
     + \frac{eV_d}{\pi} T_d\Delta\phi_t \;.
\end{equation}
Here $\Delta T_d\equiv |t_0|^2-|t_1|^2=|r_1|^2-|r_0|^2$
represents the change in the transmission probability, and
$\Delta\phi=\Delta\phi_t-\Delta\phi_r$ with $\Delta\phi_t$
($\Delta\phi_r$) being the change in the transmission
(reflection) amplitude due to different charge states:
$\Delta\phi_t=\phi_{t0}-\phi_{t1}$,
$\Delta\phi_r=\phi_{r0}-\phi_{r1}$.
Note that
$\Gamma_\phi$ was not taken into account in the analysis of the
experiment~\cite{kalish04}.
%

Given the reduced density matrix, we can evaluate the retarded
Green's function for the QD by using the formula
\begin{equation}
 G_d(\omega) = -i\int_0^\infty dt\,e^{i\omega t} \mathsf{Tr}\left(
  \rho(t) [d_\sigma(t),d_\sigma^\dagger]_+
  \right) \;,
\end{equation}
where $[\cdots,\cdots]_+$ denotes the anti-commutator.
The Green's function can be evaluated in a similar way to
the one in Ref.[\onlinecite{gunnarsson83}].
One has to use equations of motions for various
Green's functions and truncate higher order terms of $1/N_s$ with $N_s$
being the spin degeneracy. Neglecting incoherent background and the energy
shift $\Delta\epsilon$, we obtain
\begin{equation}
 G_d(\omega) \simeq \frac{(1-\bar{n_d})^2}{\omega-T_K} +
    \frac{\bar{n_d}(1-\bar{n_d})}{\omega-T_K + i\Gamma_d} \;,
  \label{eq:green}
\end{equation}

The effect of decoherence can be investigated by the $V_d$-dependence of the
conductance through the QD, $G(V_d)$, which is proportional to
$|G_d(\omega=0)|^2$ at zero temperature. We define the renormalized conductance
$g(V_d)$ as
\begin{equation}
 g(V_d) \equiv G(V_d)/G(V_d=0) \;.
\end{equation}
From Eq.(\ref{eq:green}) one can find
\begin{equation}
 g(V_d) = \frac{T_K^2+(1-\bar{n_d})^2\Gamma_d^2}{T_K^2+\Gamma_d^2} \;,
\end{equation}
which reduces to
\begin{equation}
 g(V_d) \simeq T_K^2/(T_K^2+\Gamma_d^2)
 \label{eq:conductance-kondo}
\end{equation}
in the Kondo limit~\cite{note}. Note that the $V_d$-dependence of
$g(V_d)$ comes through the relation~(\ref{eq:1/td}).

From our results
(Eqs.(\ref{eq:lambda},\ref{eq:1/td},\ref{eq:1/td-weak},
\ref{eq:conductance-kondo})) it is obvious that the Kondo
resonance is reduced by the measurement via the coherent
scattering at the QPC. This coherent scattering is described by
the complex transmission and reflection coefficients. In the weak
measurement limit, the decoherence rate is given by the
phase-sensitive ($\Delta\phi$) as well as the magnitude-sensitive
($\Delta T_d$) detection. The analysis of the
experiment~(\onlinecite{kalish04}) is based only on the first
term of Eq.(\ref{eq:1/td}). In the following, we argue that the
phase-sensitive term can be dominant in certain situation, which
may explain anomalous features reported in the experiment.

Much stronger decoherence rate than predicted in the theory based
only on $\Gamma_T$ suggests $\Gamma_\phi\gg\Gamma_T$. Here we
discuss on the condition this relation can be achieved. First,
$\Delta\phi=0$ if the QPC potential and its variation due to an
extra QD electron have mirror
symmetries~\cite{korotkov01,pilgram02}, and thus the phase-sensitive
detection does not appear.
However, in reality, there is no reason to believe that the
response of the QPC potential to the QD charge should be
symmetric. In order to consider a generic situation we introduce
asymmetric as well as symmetric variation of the QPC potential.
The model for the QPC potential is given in the form
\begin{equation}
 V(x) = \left\{ \begin{array}{ll}
    V_0 - \frac{1}{2}m\omega_x^2x^2 & \mbox{\rm for }
        -x_0 < x < x_0, \\
    \tilde{V}_0 & \mbox{otherwise} \,,
                \end{array} \right.
 \label{eq:qpc-potential}
\end{equation}
where $x_0\equiv \sqrt{\hbar/m\omega_x}$ characterizes the length
scale of the QPC with the constant potential $\tilde{V}_0=
V_0-\frac{1}{2}m\omega_x^2x_0^2$ outside the QPC region. The
potential variation due to an excess QD charge is introduced in
the following form including asymmetry,
\begin{equation}
 \delta V(x) = \left\{ \begin{array}{ll}
      \delta V_s & \mbox{\rm for }
        -a < x < x_0, \\
      \delta V_s + \delta V_a & \mbox{\rm for }
        x_0 < x < a, \\
      0 & \mbox{otherwise} \,,
                \end{array} \right.
\end{equation}
where $a$ sets the scale of the range for the potential being
affected by the excess QD charge. $\delta V_s$ and $\delta V_a$
represent the symmetric and asymmetric components of the potential
variation due to the excess QD charge, respectively. It should be
noted that $a\gg x_0$ is expected in typical geometries of
controlled dephasing
experiments~\cite{buks98,sprinzak00,kalish04}, because the QPC is
made very narrow, and the excess charge of the 
QD affects the potential in much wider region.

The transmission probability $T_d$ for the
potential~(\ref{eq:qpc-potential}) is mainly determined by the
region of $|x|\lesssim x_0$, then $T_d$ can be obtained from
the inverse harmonic potential~\cite{fertig87}
\begin{equation}
 T_d \simeq \frac{1}{1+\exp{(-\pi\varepsilon_0)}} ,
\end{equation}
where the dimensionless variable $\varepsilon_0$ is defined by
\begin{equation}
 \varepsilon_0 = 2(E-V_0)/\hbar\omega_x \;.
\end{equation}
We can obtain the change of the transmission
probability
\begin{equation}
 \Delta T_d \simeq -2\pi T_d(1-T_d)\delta v_s ,
  \label{eq:deltaTd}
\end{equation}
where $\delta v_s\equiv \delta V_s/\hbar\omega_x$. The phase
change can be estimated with the help of the WKB approximation as
\begin{equation}
 \Delta\phi \simeq \int_{x_0}^a\delta k(x)\,dx \;,
  \label{eq:deltaphi}
\end{equation}
where $\delta
k(x)\equiv\frac{\sqrt{2m}}{\hbar}[\sqrt{E-(V(x)+\Delta V_s+\delta
V_a)} - \sqrt{E-(V(x)+\Delta V_s)}]$ for $x_0<x<a$. Note that the
symmetric part $\delta V_s$ does not contribute to $\Delta\phi$
because the phase change for the transmission and the reflection
components are exactly canceled for the symmetric potential
variation~\cite{korotkov01,pilgram02}. Then we find for $a\gg
x_0$,
\begin{equation}
 \Delta\phi \simeq -\delta v_a \frac{a}{x_0} \;,
\end{equation}
where $\delta v_a\equiv \delta V_a/\hbar\omega_x$. With the help
of Eqs.~(\ref{eq:1/td-weak},\ref{eq:deltaTd},\ref{eq:deltaphi})
we obtain
\begin{subequations}
 \label{eq:Gamma-model}
\begin{eqnarray}
 \Gamma_T &=& \pi^2\frac{eV_d}{h} T_d(1-T_d)(\delta v_s)^2 \\
 \Gamma_\phi &=& \frac{eV_d}{h} T_d(1-T_d)(\delta v_a)^2 (a/x_0)^2.
\end{eqnarray}
\end{subequations}
Because $a\gg x_0$ for the typical geometry of the QPC detector,
the condition $\Gamma_\phi\gg\Gamma_T$ can be achieved if the
asymmetric component of $\delta V(x)$ is not negligible compared
to the symmetric part (That is $\delta v_a \sim \delta v_s$). The
crucial point is that the transmission probability is affected
only by the region $|x|\lesssim x_0$, while the phase is affected
through relatively wide region (up to $|x|\sim a$), so that the
phase-sensitive detection is much more effective. This can be a
natural explanation for the anomalously large decoherence rate
observed in Ref.[\onlinecite{kalish04}].


Finally, we would like to remark on the $T_d$-dependence of
$\Gamma_d$. According to Eq.(\ref{eq:Gamma-model}), both
$\Gamma_T$ and $\Gamma_\phi$ (therefore $\Gamma_d$) are
proportional to the partition noise ($\propto T_d(1-T_d)$) of the
ideal single-channel QPC. However, the experimental $\Gamma_d$-$T_d$
curve shows a double peak behavior~\cite{kalish04} in contrast to
Eq.(\ref{eq:Gamma-model}). This qualitative discrepancy might be
related to the so called ``0.7 anomaly"~\cite{0.7anomaly} where
the shot noise is also suppressed~\cite{roche04}, or to the charge
screening effect~\cite{buttiker00,buttiker03}. This issue
requires more careful experimental and theoretical analysis on the
correlation between the decoherence at the QD and the shot noise
at the QPC.



In conclusion, we have presented a theory of the controlled
dephasing of the Kondo state in a QD via the charge state
measurement in a nearby QPC device. The Kondo-assisted transport
is suppressed by the magnitude- and the phase-sensitive detection
in the QPC.  We have discussed on the condition in which the
phase-sensitive detection can be dominant in generic situation,
which may be the explanation on the unusually large decoherence
observed in a recent experiment~\cite{kalish04}. We have also
pointed out that it is important to study the correlation between
the shot noise in the detector and the decoherence in the QD.

\acknowledgements%
Helpful discussions with Y.-C.~Chung and G.~L.~Khym are greatly
appreciated. This work was supported by the KOSEF
(R05-2004-000-10826-0)

%


\begin{thebibliography}{99}
\bibitem{note} Note that the reduction factor of the conductance
 obtained here has different functional form compared with the one obtained
 from the slave-boson mean-field theory~\cite{silva03}.
\end{thebibliography}
\end{document}